\documentclass[12pt,aps,prd,preprint,nofootinbib,superscriptaddress]{revtex4-1}
\usepackage{epsf}
\usepackage{epsfig}
\usepackage{graphics,slashed}
\usepackage{graphicx}
\usepackage{amssymb}
\usepackage{amsmath}

\newcommand{\nc}{\newcommand}
\nc{\postscript}[2] 
{\setlength{\epsfxsize}{#2\hsize}\centerline{\epsfbox{#1}}}
\nc{\non}{\nonumber}
\nc{\hc}{\hbox {h.c.}} \nc{\re}{\hbox {Re}} \def\im{{\rm Im}}
\nc{\mev}{\hbox {MeV}} \nc{\gev}{\;\hbox {GeV}} \nc{\tev}{\;\hbox {TeV}}
\def\lsim{\mathrel{\raise.3ex\hbox{$<$\kern-.75em\lower1ex\hbox{$\sim$}}}}
\def\gsim{\mathrel{\raise.3ex\hbox{$>$\kern-.75em\lower1ex\hbox{$\sim$}}}}

\nc{\etal}{{\it et al.}}
\nc{\Lsp}{\;\;\;\;\;\;\;\;\;\;}  \nc{\LLLsp}{\lspace \lspace}
\nc{\lsp}{\;\;\;\;\;\;}
\nc{\spac}{\;\;\;}
\nc{\noi}{\noindent}
\nc{\beq}{\begin{equation}}   \nc{\eeq}{\end{equation}}
\nc{\bea}{\begin{eqnarray}}   \nc{\eea}{\end{eqnarray}}
\nc{\baa}{\begin{array}}      \nc{\eaa}{\end{array}}
\nc{\bit}{\begin{itemize}}    \nc{\eit}{\end{itemize}}
\nc{\ben}{\begin{enumerate}}  \nc{\een}{\end{enumerate}}
\nc{\bce}{\begin{center}}     \nc{\ece}{\end{center}}

\def\sq2{\sqrt{2}}

\def\ph{\varphi}

\def\m4{m^4(\ph)}
\def\mn2{m_n^2}

\def\v5{V^{(5)}}

\begin{document}

\title{\begin{flushright}
       \mbox{\normalsize \rm DAWSON-HEP 5, CUMQ/HEP 193}
       \end{flushright}
       \vskip 15pt
       Top and bottom partners, Higgs boson on the brane, and the tth signal
}\author{Gilles
  Couture\footnote{Email: couture.gilles@uqam.ca}} 
\affiliation{
GPTP, D\'epartement des Sciences de la Terre et de L'Atmosph\`ere,
Universit\'e du Qu\'ebec \`a Montr\'eal, Case Postale 8888, Succ. Centre-Ville,
 Montr\'eal, Qu\'ebec, Canada, H3C 3P8}
\author{Mariana Frank\footnote{Email: mariana.frank@concordia.ca}}
\affiliation{Department of Physics, Concordia University,
  7141 Sherbrooke St. West, Montreal, Quebec, CANADA H4B 1R6}
\author{Cherif Hamzaoui\footnote{Email: hamzaoui.cherif@uqam.ca}}
\affiliation{
GPTP, D\'epartement des Sciences de la Terre et de L'Atmosph\`ere,
Universit\'e du Qu\'ebec \`a Montr\'eal, Case Postale 8888, Succ. Centre-Ville,
 Montr\'eal, Qu\'ebec, Canada, H3C 3P8}
\author{Manuel Toharia\footnote{Email: mtoharia@dawsoncollege.qc.ca}}
\affiliation{ Physics Department, Dawson College,
 3040 Sherbrooke St., Westmount, Quebec, Canada H3Z 1A4}
\affiliation{Department of Physics, Concordia University,
  7141 Sherbrooke St. West, Montreal, Quebec, CANADA H4B 1R6}

\date{\today}

\begin{abstract}
  Current LHC results indicate a possible enhancement in the production of Higgs bosons in
   association with top quarks ($t\bar{t}h$) over the Standard Model (SM) expectations, suggesting an
   increase in the top Yukawa coupling. To explain these results, 
   we study the effect of adding to the SM a small set of vector-like
   partners of the top and  bottom quarks with masses of order $\sim 1$ TeV.
   We consider Yukawa coupling matrices with vanishing
   determinant and show that then, Higgs production through gluon
   fusion is not affected by deviations in the top quark Yukawa
   coupling, and in fact depends only on deviations in the bottom quark
   Yukawa coupling. We call this scenario the {\it Brane Higgs Limit},
   as it can emerge naturally in models of warped extra-dimensions
   with all matter fields in  the bulk, except the Higgs (although it could also occur in 4D
   scenarios with vector-like quarks and special flavor symmetries
   forcing the vanishing of the Yukawa determinants). We show that the scenario
   is highly predictive for all Higgs production/decay modes, making it
   easily falsifiable, maybe even at the LHC RUN 2 with higher luminosity.

\end{abstract}

\maketitle

\section{Introduction}
\label{sec:intro}  

RUN 1 of the LHC culminated in the discovery of the Higgs boson at 125
GeV.  After the Higgs discovery, the most important question is,
naturally, where is the new physics beyond the Standard Model (SM),
and how will it manifest itself? Minimal additions to the SM could be revealed  as an
unexpected excess in dileptons, $W^+W^-$,  $ZZ$, diphotons, $t\bar{t}$
or  $b \bar{b}$, indicating the presence of a new boson
resonance. More involved new physics could also appear as missing
energy, or any other signals indicating the presence of 
supersymmetry or extra-dimensions or other exotic particles. But also, new physics could
manifest itself indirectly. In particular it could affect the production
cross section and decay widths of the Higgs boson, expected to be
measured with increased precision during the current RUN2 of the LHC. There are already 
several promising signals in the RUN 1 data indicating possible
deviations from the SM expectations. In particular, both ATLAS and CMS
report a possible increase in the signal strength of the $t{\bar t}h$
associated production in the LHC data. Of particular interest
from RUN 1 at CMS and ATLAS are the same-sign dilepton (SS2$l$) and
trilepton ($3l$) signals coming from leptonic Higgs decays in the associated 
$t{\bar t}h$ production events.
The best fit signal strengths are found to be
$\mu_{SS2l}=5.3^{+2.1}_{-1.8}$ and  $\mu_{3l}= 3.1^{+2.4}_{-2.0}$ at
CMS \cite{Khachatryan:2014qaa}, and  $\mu_{SS2l}=2.8^{+2.1}_{-1.9}$  and
$\mu_{3l}= 2.8^{+2.2}_{-1.8}$ at ATLAS \cite{Aad:2015iha}.
These leptonic excesses  are associated to the channels $tt(h\to
WW^*)$, $tt(h\to ZZ^*)$ and $tt(h\to \tau \tau)$ where one of the tops
decays leptonically.
Within the preliminary results of RUN 2 in those same leptonic channels, both ATLAS and CMS still
report excesses with, for example, $\mu_{SS2l}=1.9^{+0.9}_{-0.8}$ at CMS
\cite{CMS:2016vqb} and  $\mu_{SS2l}=4.0^{+2.1}_{-1.7}$ at
ATLAS \cite{ATLAS:2016awy}.
The most recent preliminary results reported by CMS in the $t{\bar t}h$
associated production searches make use of an integrated
luminosity of 35.9 fb$^{-1}$ and seem to show mixed results. In the leptonic
channels ($W^+W^-$ and $ZZ$ channels) they still show an enhancement of $1.5 \pm0.5$
times the SM prediction, with an observed (expected) significance of
3.3 $\sigma$ (2.5 $\sigma$) obtained  from combining these results
with the 2015 data \cite{CMS:2017vru}. On the other hand in the
$h\to\tau\tau$ decay channel search, a slight suppression of $0.72^{+0.62}_{-0.53}$
times the SM prediction is found, with an observed (expected) significance of
$1.4\sigma$ ($1.8\sigma$) \cite{CMS:2017lgc}. Note that, unlike the $W^+W^-$ and $ZZ$ decays, this last
signal is sensitive to both the top-Higgs Yukawa and 
the $\tau$-Higgs Yukawa couplings, and thus enhancements or suppressions  are possible  as long as there are 
variations in {\it either} the top quark and the $\tau$ lepton Yukawa couplings.      

All measurements are still hindered by having few events so
far, but 
nevertheless, should these tantalizing signals survive more precise measurements at higher
luminosities, they will provide the much awaited signals for new
physics. We summarize relevant production and decay channels in Table
\ref{tab:HIGGS_ttH_12} with the overall combinations obtained by the ATLAS
and CMS collaborations, for the  signal strengths associated to
each Higgs production and decay channels.

\begin{table}[tb]
\begin{center}
\begin{tabular}{|c|c|c|c|c|c|c|}
\hline
Production Mode & Channel   &   RUN-1 \cite{Khachatryan:2016vau} &  Production Mode & Channel & ATLAS RUN-2  &  CMS RUN-2 \\   
\hline
\hline
$ggh$ & $\gamma \gamma$  & $1.1^{+0.23}_{-0.22}$& $ggh$ & $\gamma \gamma$  & $0.62^{+0.30}_{-0.29}$  \cite{ATLAS:2016hru} &   $0.77^{+0.25}_{-0.23}$  \cite{CMS:2016ixj}
\\ \cline{2-5}  
&$WW^*$   &  $0.84^{+0.17}_{-0.17} $   
&&$WW^*$  & - & - 
\\ \cline{2-5} 
& $ZZ^*$   &  $1.13^{+0.34}_{-0.31} $  & 
& $ZZ^*$   & $1.34^{+0.39}_{-0.33} $ \cite{ATLAS:2016hru} & $0.96^{+0.44}_{-0.33} $ \cite{CMS:2016ilx}
\\ \hline
$ t{\bar t}h$ &$\gamma \gamma$  & $2.2^{+1.6}_{-1.3}$& $ t{\bar t}h$ &$\gamma \gamma, b{\bar b}, $ leptons  & $1.8^{+0.7}_{-0.7}$  \cite{ATLAS:2016axz} &   - \\ \hline 
  & $b {\bar b}$& $1.15^{+0.99}_{-0.94}$& &$WW^\star ,ZZ^\star, \tau^+\tau^-$  &  &  $2.0^{+0.8}_{-0.7}$ \cite{CMS:2016vqb} \\ \hline 
 & $W W^\star$& $5.0^{+2.6}_{-2.2}$& &   & & 
\\ \hline 
\end{tabular}
\end{center}
\caption{
\label{tab:HIGGS_ttH_12}
Higgs signal strengths used in our analysis, from $ggh$ and $ t{\bar t}h$   
production modes measured at the LHC from RUN 1 (combined $\sqrt{s}=7$
and $8$ TeV results)  and RUN 2 ($\sqrt{s}=13$  TeV).  
}
\end{table}

One possibility to explain the SS2$l$ excess,   
is that it  could be due to a modified Higgs coupling to the SM
top quark, resulting in an enhanced $t{\bar t}$ ($h \to$ multileptons)
production. A simple explanation put forward to explain this latest
possible signal of physics beyond the SM has been to
invoke the presence of vector-like quarks
\cite{Angelescu:2015kga}. Previous studies have adopted 
an effective theory approach, involving generic couplings and mixings
with the third generation quarks, by which they induce modifications
of the Yukawa couplings of the top and bottom quarks. The scenario has
been put forward to explain deviations from SM expectations in the
forward-backward asymmetry in $b$ decays $A_{\rm FB}^b$ and the
enhancement of the $pp \to t {\bar t}h$ cross section at the
LHC. Mixing with the additional states in the bottom sector allows for
a sufficiently large increase of the $Zb_R{\bar b}_R$ coupling to
explain the forward-backward anomaly, as well as imply new effects in
Higgs phenomenology  \cite{Choudhury:2001hs,vectorlike}. 
The mixing could provide a strong enhancement of the $t{\bar t}h$
Yukawa coupling, which would explain an increase of the cross section
at the LHC. In this scenario, rates for the loop-induced processes
stay SM-like due to either small vector-like contributions or
compensating effects  between fermion mixing and loop
contributions. For this to be a viable scenario, vector-like quark
masses of order 1-2 TeV are required, still safe from the LHC lower
limits on their masses, $m_{\rm VLQ} \gsim 800$ GeV
\cite{Aad:2015mba}.\footnote{Alternative explanations involving
  supersymmetric partners have also been put forward
  \cite{Huang:2015fba}, as well as early studies on the implications
  on some coefficients of operators of the effective lagrangian
  \cite{Okada}.}   

In Section \ref{sec:!D1S} of this article, we first revisit the SM augmented by the addition of
one vector-like quark doublet and two singlets (one top-like and one
bottom-like) and review the mixing with the third family of quarks.
In Sec. \ref{subsec:braneHiggs}, we then show that there is a specific
region of parameter space where large corrections to the top Yukawa
coupling (caused by contributions from the new vector-like quarks) do not cause large
corrections to the radiative coupling of the Higgs and gluons.
We show that in that limit, Higgs signal strengths can be simply
parametrized in terms of four variables only, related to the top and
bottom quark shifts in Yukawa couplings, which we  analyze in
Sec. \ref{subsec:Yukawas}, in terms of the parameters of the model.
Based on these results, we present simple predictions between signal strength in
Higgs production (through gluon fusion and $t {\bar t}h$) and decays
into $\gamma \gamma,\, WW$ and $ZZ$, in Sec. \ref{subsec:pheno}. As seen in
that subsection, large regions of parameter space are excluded by both
theoretical considerations and experimental constraints. Finally, in
Sec \ref{sec:RS} we describe how to reproduce our scenario within the
conventional Randall Sundrum model and then summarize our findings in
Sec. \ref{sec:conclusion}.

\section{Top and Bottom Mirrors: a doublet and two singlets}
\label{sec:!D1S}

The simple scenario that we wish to consider contains the usual SM
gauge groups and matter fields, with the addition of a 
vector-like quark $SU(2)$ doublet and two vector-like quarks
$SU(2)$ singlets, one with up-type gauge charge and another with
down-type gauge charge. They can be regarded 
as top or bottom partners as we will consider that their  Yukawa couplings
are large. As we will show in the next section, this structure can appear naturally in models of warped
extra dimensions with the Higgs localized near the TeV brane, and with
fermions in the bulk. The presence of brane kinetic terms can lower
significantly the mass of some of the heavy Kaluza Klein fermions \cite{Frank:2016vtv}. The
rest of the KK fields decouple due to their heavy masses , giving rise
to something similar to the simple setup considered here. 
We denote
$ q^0_L\equiv \begin{pmatrix}
  t^0_L  \\
  b^0_L
\end{pmatrix} $
as the SM third generation doublet,  
and $t^0_R$ as the SM right handed top. Using similar notation we 
define $Q_{L,R}\equiv\begin{pmatrix}
  Q^t_{L,R}  \\
  Q^b_{L,R}
\end{pmatrix}$
as the new vector-like quark doublet,  $T_{R,L}$ as the new
vector-like up-type quark singlet, and $B_{L,R}$ as the  new
vector-like down-type singlet.

In principle we should also consider the mixings with the $up$ and
$charm$ quarks, and $down$ and $strange$ quarks of the SM when writing down the most general Yukawa
couplings in the up and down sectors.
Without any additional assumption or theory input, we should write down the most general Yukawa
couplings between SM quarks and the new vector-like quarks, leading to a
$5\times5$ fermion mass matrices. In models of warped extra dimensions
with bulk fermions, the couplings between  up or charm and heavy fermions are suppressed by factors of
  order $\sqrt{m_f/m_t}$ with respect to the couplings to the top,
  which are of order 1, and so we  will just neglect those terms, leading to
  a much simpler $3 \times 3$ fermion mass matrix (and similarly in
  the down sector with the bottom quark).
  
  The mass and interaction Lagrangian in the top sector, including
  its Yukawa couplings with the SM-like Higgs doublet $\tilde{H}$ can be then written as
\bea
{\cal L}_{mass}&=& Y_t^0\ \overline{q}_L\tilde{H} t_R +Y_{qT}\  \overline{q}_L\tilde{H}T_R
+Y_{Qt}\ \overline{Q}_L\tilde{H}t_R  + Y_{1}   \overline{Q}_L\tilde{H} T_R +
Y_{2}   \overline{Q}_R \tilde{H} T_L   \non\\
&& + M_Q \overline{Q}_L Q_R + M_T \overline{T}_LT_R,
\eea
with a similar expression for the bottom sector.
After electroweak symmetry breaking, the Yukawa couplings induce off-diagonal terms
into the fermion mass matrix. In the basis defined
by the vectors $(\overline{q}_L,\overline{Q}_L,\overline{T}_L)$ and $(t_R,Q_R,T_R)$
we can write the heavy quarks mass matrix as
\bea
{\bf M_t}=\begin{pmatrix}
v Y_t^0 &0& v Y_{qT}\\
v Y_{Qt}&M_Q& vY^t_{1}\\
0&vY^t_{2} &M_T
\end{pmatrix}.\label{Mt}
\eea
where, in general, all entries are complex\footnote{We can eliminate
  five phases from the mass matrix ${\bf M_t}$ through phase redefinitions. We 
  keep the notation general, since the phases regroup together easily
  in all the expressions.} 
and where $v$ is the Higgs vacuum expectation value (VEV). 
We can also express the bottom sector heavy quark mass matrix as
\bea
{\bf M_b}=\begin{pmatrix}
vY_b^0 &0& vY_{qB}\\
vY_{Qb}&M_Q& vY^b_{1}\\
0&vY^b_{2}&M_B
\end{pmatrix}~,
\label{Mb}
\eea
where again all entries can be complex and the value of $M_Q$ is the
same in both ${\bf M_t}$ and ${\bf M_b}$.
The associated top and bottom Yukawa coupling matrices are 
\bea
{\bf \tilde{Y}_t}=\begin{pmatrix}
Y_t^0 &0& Y_{qT} \\
Y_{Qt}  &0& Y^t_{1}  \\
0&Y^t_{2}  &0
\end{pmatrix}\ \ \ {\rm and}\  \  \ \  {\bf \tilde{Y}_b}=\begin{pmatrix}
Y_b^0 &0& Y_{qB} \\
Y_{Qb}  &0& Y^b_{1}  \\
0&Y^b_{2} &0
\end{pmatrix}. \label{Yt}
\eea
The mass matrices ${\bf M_t}$  and ${\bf M_b}$ are diagonalized by bi-unitary transformations,
${\bf V_t}^\dagger_L{\bf M_t}{\bf V_t}_R={\bf M_t}^{diag}$,  and
${\bf V_b}^\dagger_L{\bf M_b} {\bf V_b}_R={\bf M_b}^{diag}$.
At the same time, the Higgs Yukawa couplings are obtained after
transforming the Yukawa matrices into the physical basis,
${\bf V_t}^\dagger_L {\bf \tilde{Y}_t}{\bf V_t}_R={\bf Y_t}^{phys}$
and ${\bf V_b}^\dagger_L {\bf \tilde{Y}_b}{\bf V_b}_R={\bf Y_b}^{phys}$.

\subsection{Higgs Production in the {\it Brane Higgs Limit}}
\label{subsec:braneHiggs}

In the physical basis, the top quark mass and the top Yukawa coupling
(the first entries in the physical mass matrix and the physical Yukawa
matrix) are not related anymore by the SM relationship $m_t^{phys}= v
y_t^{SM}$  \cite{Azatov:2009na} (with $v$ normalized to $v=174$ GeV for simplicity). 
The same goes for the bottom quark, and we thus define the 
shifts, $\delta y_t$ and $\delta y_t$, between the SM and the physical
Yukawa couplings, due to the diagonalization, as
\bea
\label{deltayt}
y^{phys}_t = y^{SM}_t - \delta y_t
\eea
and
\bea
\label{deltayb}
y^{phys}_b = y^{SM}_b - \delta y_b.
\eea
Later  we will give an approximate expression of these shifts in terms
of the model parameters of Eqs.~(\ref{Mt}) and (\ref{Mb}). But before that, we
consider the radiative coupling of the Higgs to gluons. This coupling
 depends on the physical Yukawa couplings $y_{nn}$ of all
the fermions running in the loop and on their physical masses
$m_n$. The real and imaginary parts of the couplings (the scalar  and
pseudoscalar parts) contribute to the cross section through different loop functions,
$A^S_{1/2}$ and $A^P_{1/2}$, as they generate the two operators $h
G_{\mu\nu}G^{\mu\nu}$ and $h G_{\mu\nu}\tilde{G}^{\mu\nu}$. 

\begin{figure}[t]
\center
\begin{center}
	\includegraphics[height=4.cm]{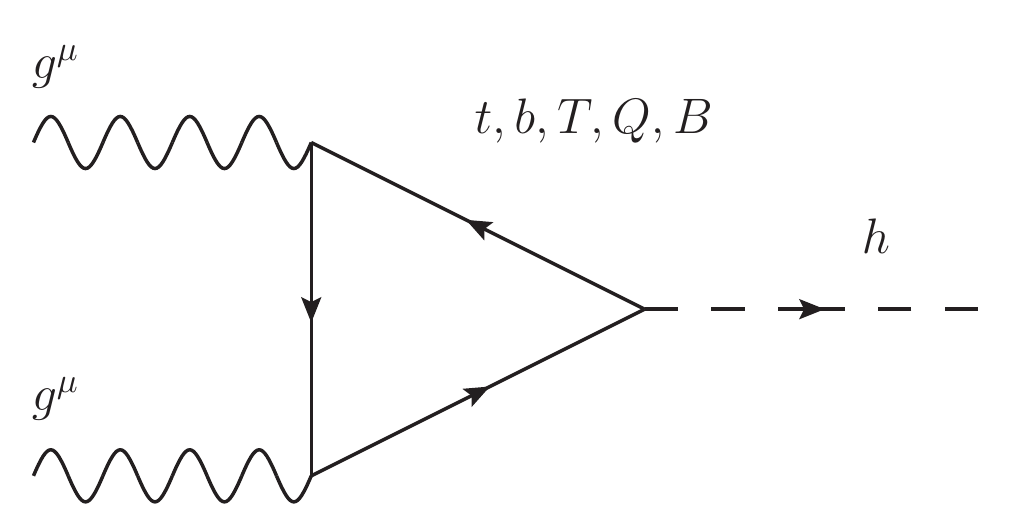}
 \end{center}
\vspace{.1cm}
\caption{Feynman diagram for the production cross section $gg \to h$
  in a setup with new vector-like fermions $Q$, $T$ and $B$.
 }  \label{fig:decayhgg}
\vspace{.4cm}
\end{figure}
The cross section, depicted in Fig. \ref{fig:decayhgg}, is 
\bea
\sigma_{gg\rightarrow h} = {\alpha_s^2 m_h^2\over 576 \pi} \left[|c^S_{ggh}|^2 + |c^P_{ggh}|^2\right]\ \delta(s-m_h^2)
\eea
where
\bea
 c^S_{ggh} =\sum^3_{n=1} \re\left({\frac{y_{nn}}{m_{n}}}\right) A^S_{1/2}(\tau_f)
\hspace{.6cm} {\rm and} \hspace{.6cm}
c^P_{ggh} = \sum^3_{n=1} \im\left(\frac{y_{nn}}{m_{n}}\right) A^P_{1/2}(\tau_f)
\eea
with $\ \tau = m^2_h/4m^2_n\ $ and with the loop functions
$A^S_{1/2}(\tau)$ and  $A^P_{1/2}(\tau)$ as defined in \cite{Gunion:1989we}.
Note that we use a normalization  of the loop functions such that for very heavy quarks
with masses $m_n$ much greater than the Higgs mass 
$m_h$ (i.e. when $\tau$ is very small) they behave
asymptotically as $\displaystyle \lim_{\tau 
  \to 0} A^S_{1/2} =1 \hspace{.2cm} \mbox{and} \hspace{.2cm}
\lim_{\tau \to 0} A^P_{1/2} =3/2.$ On the other hand, for light quarks (all the SM quarks
except top and to some extent, bottom), the loop functions essentially vanish 
since $\displaystyle \lim_{\tau \to \infty} A^S_{1/2}
= \lim_{\tau \to \infty} A^P_{1/2} = 0$, and we thus neglect
contributions from the light SM quarks and consider only the
effect of the top, the bottom and the four remaining  physical heavy quarks.  

The amplitudes $c^S_{ggh}$ and $c^P_{ggh}$
can then be written in terms of traces involving the fermion mass and
Yukawa matrices involving top  and vector-like up-quarks, and
bottom  and any vector-like down quarks, ${\bf M}_i$ and ${\bf
  Y}_i$ with $i=t,b$, so that we obtain
\bea
c^S_{ggh} = \sum_n \re\left({\frac{y^u_{nn}}{m^u_{n}}}\right) +
\sum_n \re\left({\frac{y^d_{nn}}{m^d_{n}}}\right ) -
\re\left({\frac{y_b}{m_b}} \right)+  \re\left({\frac{y_b}{m_b}}\right) A^S_{1/2}(\tau_b)
\label{csggh}
\eea
where we have added and subtracted the bottom quark loop contribution
in order to keep the dependence in $A_{1/2}(\tau_b)$, and with a similar expression holding for $c^P_{ggh}$.
We evaluate exactly the sums in the top and bottom sectors and find
\bea
\sum_n \left({\frac{y^u_{nn}}{m^u_{n}}}\right)&=&
\frac{1}{v} \frac{1 + 3 \varepsilon_{Q_t} \varepsilon_T
  \frac{|Y^t_2|}{|Y_t^0|}e^{i\theta^t_2}\left(1
  -e^{i\theta^t_1}\frac{|Y^t_1||Y_t^0|}{|Y_{Qt}||Y_{qT}|}
  \right)}{1 + \varepsilon_{Q_t} \varepsilon_T
  \frac{|Y^t_2|}{|Y_t^0|}e^{i\theta^t_2}\left(1
  -e^{i\theta^t_1}\frac{|Y^t_1||Y_t^0|}{|Y_{Qt}||Y_{qT}|}
  \right)}, \label{hggtopcontribution}
\eea
and
\bea
\sum_n \left({\frac{y^d_{nn}}{m^d_{n}}}\right)&=&
\frac{1}{v} \frac{1 + 3 \varepsilon_{Q_b} \varepsilon_B
  \frac{|Y^b_2|}{|Y_b^0|}e^{i\theta^b_2}\left(1
  -e^{i\theta^b_1}\frac{|Y^b_1||Y_b^0|}{|Y_{Qb}||Y_{qB}|}
  \right)}{1 + \varepsilon_{Q_b} \varepsilon_B
  \frac{|Y^b_2|}{|Y_b^0|}e^{i\theta^b_2}\left(1
  -e^{i\theta^b_1}\frac{|Y^b_1||Y_b^0|}{|Y_{Qb}||Y_{qB}|}
  \right)}, \label{hggbottomcontribution}
\eea
where we have defined the small parameters
$\displaystyle \varepsilon_T=\frac{v| Y_{qT}|}{|M_T|}\ $,
$\displaystyle \varepsilon_B=\frac{v| Y_{qB}|}{|M_B|}\ $,
$\displaystyle\ \varepsilon_{Q_t}=\frac{v|Y_{Qt}|}{|M_Q|}\ $ and
$\displaystyle\ \varepsilon_{Q_b}=\frac{v|Y_{Qb}|}{|M_Q|}\ $, and with
the relative phases $\theta^i_1$ and $\theta^i_2$ defined as
$\theta^t_1 = Arg\left(\frac{Y_t^0 Y^t_1}{Y_{Qt}Y_{qT}} \right)$ 
and $\theta^t_2 = Arg\left(\frac{Y_{qT}Y_{Qt} Y^t_2}{M_TM_QY^0_t}
\right)$, with similar definitions for $\theta_1^b$ and
$\theta_2^b$. In the SM limit, the expression in Eq. (\ref{csggh}) should tend to 
$\sim \displaystyle \frac{1}{v}(1+A_{1/2}^S(\tau_b))$, and so if one
wished to limit the contribution coming from the top partners to
Higgs production (in gluon fusion), we must 
reduce/eliminate these corrections.  We note the following observations.
\bit
\item Of course for heavier and heavier vector-like fermions, the parameters $\varepsilon_i$
become more and more suppressed, and thus we can smoothly recover the SM
limit, but the new physics effect will decouple from everywhere else (and in particular
the top Yukawa quark will also tend to its SM value).
\item It might also be possible to reduce the couplings $|Y^t_2|$ or $|Y^b_2|$ , but
then this will also affect the physical top or bottom Yukawa couplings shifts,
and in particular no enhancement in the top quark Yukawa coupling
will be possible (although suppression might still be possible), as we
will show later.

\item Another interesting possibility would be to set the overall phase of the
correction term in Eqs.~(\ref{hggtopcontribution}) or
(\ref{hggbottomcontribution}) to be $\pi/2$ so that the real part
vanishes (in general, we expect that the real part would dominate the overall corrections, at
least for $\varepsilon_i <1$). This possibility might limit the amount of enhancement in
the top Yukawa 
coupling,  since that  correction also depends on the phase $\theta_2$. Again
when we compute the approximate expression  of the Yukawa couplings shift, we
will see that the phase $\theta_2$  should be close to $0$ to yield  an
enhancement in the top Yukawa coupling. 

\item In our considerations, we will impose a seemingly contrived constraint
  on the model parameters, which we call the {\it Brane Higgs Limit},
  such that
  \bea
  \det {\bf \tilde{Y}_t} =  \det {\bf \tilde{Y}_b} =0,
  \eea
with the matrices ${\bf \tilde{Y}_t}$ and ${\bf \tilde{Y}_b}$ defined in Eq.~(\ref{Yt}). This
constraint implies that  
\bea
Y^t_2 \left(1 -e^{i\theta_1}\frac{|Y^t_1||Y_t^0|}{|Y_{Qt}||Y_{qT}|}\right)=0, \label{cancellation}
\eea
and thus ensures that the top sector contribution to Higgs production,
given in Eq.~(\ref{hggtopcontribution}), gives the same result as the SM
top quark contribution to the same process.  The vanishing determinant
condition  could come from a specific flavor structure in  the Yukawa
matrix, emerging for example from democratic  textures, etc.  We will
show in the next section that the flavor   structure required can also be obtained in models of
  extra-dimensions, so that the cancellation in
  Eq.~(\ref{cancellation}) is satisfied exactly if the scenario arises out of
  the usual Randall-Sundrum warped extra-dimensional scenario with
  matter fields in bulk. It is necessary, though, that the Higgs be
  sufficiently localized  towards the brane and that the KK modes of the
  top quark (and bottom quark) be much lighter than the KK partners
  of the up and charm quarks (and the down and strange quarks). We will then refer to the vector-like partners of the top and bottom quarks throughout as KK partners, and we return to this scenario in Sec. \ref{sec:RS}.

\eit

Therefore we work in the {\it Brane Higgs Limit} of the general parameter
space. In the down sector, we  also have
$Y^b_2
\left(1-e^{i\theta^b_1}\frac{|Y^b_1||Y_b^0|}{|Y_{Qb}||Y_{qB}|}\right)=0$,
so that 
we have
\bea 
\sum_n \left({\frac{y^u_{nn}}{m^u_{n}}}\right)\ =\ \sum_n
\left({\frac{y^d_{nn}}{m^d_{n}}}\right)\ =\ \frac{1}{v} \, . 
\eea
This means that now we can write the $ggh$ couplings as  
\bea
c^S_{ggh} &=& \frac{1}{v}\left(1+A^S_{1/2}(\tau_b)\right) + \re\left(\frac{\delta
   y_b}{m_b}\right) \left(1- A^S_{1/2}(\tau_b) \right) \, , \\
\hspace{-3cm}{\rm  and}\hspace{3cm} &&\non\\
c^P_{ggh} &=&\im\left( \frac{\delta y_b}{m_b}\right) \left(\frac{3}{2}-A^P_{1/2}(\tau_b)\right),\  \ \ \ \  \
\eea
where we have used the definitions of the Yukawa coupling shifts in Eqs.~(\ref{deltayt}) and (\ref{deltayb}).
Evaluating the values for the bottom quark loop functions
$A^{S,P}_{1/2}(\tau_b)$ we obtain
\bea 
\frac{\sigma_{gg\rightarrow h}}{\sigma^{SM}_{gg\rightarrow
  h}} =\frac{\Gamma_{h\rightarrow gg}}{\Gamma^{SM}_{h\rightarrow
  gg}}=  (1+ \Delta_{gg})~,
\eea
where the correction term is
\bea
\Delta_{gg}=2.13 v  \left( \re \frac{\delta y_b}{m_b}\right)  
  +1.13 v^2\left(\re \frac{\delta y_b}{m_b}\right)^2 +2.51
  v^2\left(\im \frac{\delta y_b}{m_b}\right)^2.
  \eea
This result links in a simple and nontrivial way  Higgs production through gluon fusion to the bottom quark
Yukawa coupling (or more precisely to its relative shift $v\delta
y_b/m_b$). In a similar fashion we can also obtain the correction to the Higgs
decay into $\gamma\gamma$ in the {\it Brane Higgs Limit}, since the fermion loop is the same as
the gluon fusion loop (although there is an additional $W$ loop contribution in this
case).
We obtain
\bea
\frac{\Gamma_{h\to \gamma\gamma}}{\Gamma^{SM}_{h\to
  \gamma\gamma}}= (1 + \Delta_{\gamma\gamma})\, ,
\eea
with
\bea
\Delta_{\gamma\gamma} &=& -0.14 v  \left( \re \frac{\delta
  y_b}{m_b}\right)  + 0.005 v^2  \left( \re
\frac{\delta y_b}{m_b}\right)^2+ 0.01 v^2  \left( \im \frac{\delta y_b}{m_b}\right)^2 , 
\eea
and where we took the SM loop contributions to be $|c_{\gamma\gamma}| = |-7 A_1(\tau_W) + 16/9
A^S_{1/2}(\tau_t) +4/9 A^S_{1/2}(\tau_b)|
\simeq 6.53$, with the W-loop function $A_1(\tau_W)$
\cite{Gunion:1989we} and the fermion loop function  $A^S_{1/2}(\tau_f)$ normalized so that
$\lim_{\tau\to 0} A_i(\tau)=1$.
Finally, from Eqs.(\ref{deltayt}) and (\ref{deltayb}) we can now write
\bea
\frac{\sigma_{pp\rightarrow tth}}{\sigma^{SM}_{pp\rightarrow tth}} =
\left|\frac{y_t}{y_{SM}}\right|^2 = (1 +\Delta_{tt})   \ \ \ \ \ \ \  {\rm 
  and} \ \ \ \  \ \ \ \frac{\Gamma_{h\to b}}{\Gamma^{SM}_{h\to
  bb}}=\left|\frac{y_b}{y_{SM}}\right|^2 = (1
+\Delta_{bb}),\ \ \ \ 
\eea
where
\bea
\Delta_{tt}= - 2 v  \re \left(\frac{\delta
  y_t}{m_t}\right)  + v^2\left|\frac{\delta y_t}{m_t}\right|^2 \ \ \ \  {\rm
  and} \ \ \ \ \Delta_{bb}= - 2v \re \left(\frac{\delta
  y_b}{m_b}\right)  +v^2\left|\frac{\delta y_b}{m_b}\right|^2 \, .
\eea
We are interested in studying the dependence on the Yukawa shifts $\delta y_t$ and
$\delta y_b$ of the signal strengths 
\bea
\displaystyle \mu^{ii}_{ggh} = 
\frac{\sigma(gg\to h) Br(h\to ii)}{\sigma_{SM}(gg\to h) Br^{SM}(h\to
  ii)}\ = \frac{\sigma(gg\to h)}{\sigma_{SM}(gg\to h)}\ \frac{\Gamma(h\to
  ii)\ }{\Gamma_{SM}(h\to ii)}\ \frac{\Gamma_{SM}^{tot}(h)}{\Gamma^{tot}(h)}, 
\eea
with a similar expression for $\mu^{ii}_{t\bar{t}h}$ (and using the small
width approximation).
In these expressions, the ratio of total Higgs widths can be written as
\bea
\frac{\Gamma_{SM}^{tot}(h)}{\Gamma^{tot}(h)}&=& \frac{1}{\left(1+
  Br^{SM}_{h\to bb}\Delta_{bb} +
  Br^{SM}_{h\to gg}\Delta_{gg}+Br^{SM}_{h\to \gamma\gamma}\Delta_{\gamma\gamma}\right)}\ \ \ \, ,
\eea
where,  taking into account numerical values for the SM Higgs branching ratios,  gives simply
\bea
\frac{\Gamma_{SM}^{tot}(h)}{\Gamma^{tot}(h)}&\simeq& \frac{1}{1+ 0.58
  \Delta_{bb} + 0.086 \Delta_{gg}}~,
\eea
and where we have dropped the dependence in $\Delta_{\gamma\gamma}$ as it
  is much suppressed. 

  With all these ingredients, we find the $t{\bar t}h$ production-and-decay strengths
  \bea
  \mu^{VV}_{t{\bar t}h}&=& \frac{(1+\Delta_{tt})}{(1+ 0.58
  \Delta_{bb} + 0.086 \Delta_{gg})} \, , \\
  \mu^{bb}_{t{\bar t}h}&=&\mu^{VV}_{t{\bar t}h} (1+\Delta_{bb}) \, , \\
  \mu^{\gamma\gamma}_{t{\bar t}h}&=& \mu^{VV}_{t{\bar t}h} (1+\Delta_{\gamma\gamma})~, 
  \eea
  as well as the $ggh$ strengths
  \bea
  \mu^{VV}_{ggh}&=& \frac{(1+\Delta_{gg})}{(1+ 0.58
    \Delta_{bb} + 0.086 \Delta_{gg})} \, , \\
  \mu^{\gamma\gamma}_{ggh}&=& \mu^{VV}_{ggh} (1+\Delta_{\gamma\gamma})~, 
  \eea
  with the corrections $\Delta_{ii}$ depending only on top or bottom quark Yukawa coupling shifts
  \bea
  \Delta_{tt}&=& - 2 v  \re \left(\frac{\delta
  y_t}{m_t}\right)  + v^2\left|\frac{\delta y_t}{m_t}\right|^2
  \ \  \label{Dtt}\\
  \Delta_{bb} &=& - 2v \re \left(\frac{\delta
  y_b}{m_b}\right)  +v^2\left|\frac{\delta y_b}{m_b}\right|^2 \label{Dbb} \\
 \Delta_{gg}&=&2.13 v  \left( \re \frac{\delta y_b}{m_b}\right)  
  +1.13 v^2\left(\re \frac{\delta y_b}{m_b}\right)^2 +2.51
  v^2\left(\im \frac{\delta y_b}{m_b}\right)^2 \label{Dgg}\\
  \Delta_{\gamma\gamma}&=&  -0.14 v  \left( \re \frac{\delta
  y_b}{m_b}\right)  + 0.005 v^2  \left( \re
\frac{\delta y_b}{m_b}\right)^2+ 0.01 v^2  \left( \im \frac{\delta y_b}{m_b}\right)^2  \label{Dgaga}~.
  \eea


\subsection{Yukawa Coupling Shifts}
\label{subsec:Yukawas}


As indicated before, the mass matrices ${\bf M_t}$ and ${\bf M_b}$ from Eqs.~(\ref{Mt}) and (\ref{Mb})
are diagonalized by bi-unitary transformations, 
${\bf V_t}^\dagger_L{\bf M_t}{\bf V_t}_R={\bf M_t}^{diag}$,  and
${\bf V_b}^\dagger_L{\bf M_b}{\bf V_b}_R={\bf M_b}^{diag}$. In order to obtain simple
analytical expressions for the Yukawa couplings emerging after the
diagonalization, we  expand the unitary matrices  ${\bf V_{t,b}}_L$
and ${\bf V_{t,b}}_R$ in powers of $\varepsilon \sim v/M$, where $v$
is the Higgs VEV and $M$ represents the vector-like masses $M_Q$, $M_T$
or $M_B$.

In this approximation we can obtain the lightest mass eigenvalues (the top quark and
the bottom quark masses) as well as the physical $t{\bar t}h$ Yukawa coupling and
the $b{\bar b}h$ Yukawa coupling.

This yields the relative deviation between the physical
Yukawa couplings $y_t^{phys}$ and $y_b^{phys}$, and the SM Yukawa
couplings, defined as $y_t^{SM}= m_t^{phys}/v$  and $y_b^{SM}=
m_b^{phys}/v$. In terms of the mass matrix parameters from Eqs.~(\ref{Mt}) and (\ref{Mb}), we obtain
\bea
\frac{\delta y_t}{y_t^{SM}}\
=\ \varepsilon_T^2 + \varepsilon_{Q_t}^2
- 2 \varepsilon_T \varepsilon_{Q_t} \frac{|Y_2|}{|Y_t^0|}
e^{i\theta^t_2} \ +\  {\cal O}(\varepsilon^4), \label{deltayt2} 
\eea
and similarly for the bottom quark
\bea
\frac{\delta y_b}{y_b^{SM}}
=\  \varepsilon_B^2 +
\varepsilon_{Q_b}^2 - 2 \varepsilon_B \varepsilon_{Q_b} \frac{|Y^b_2|}{|Y_b^0|}
e^{i\theta^b_2} \ +\ {\cal O}(\varepsilon^4).  \label{deltayb2} 
\eea
As  previously, 
$\displaystyle \varepsilon_T=\frac{v| Y_{qT}|}{|M_T|}\ $,
$\displaystyle \varepsilon_B=\frac{v| Y_{qB}|}{|M_B|}\ $,
$\displaystyle\ \varepsilon_{Q_t}=\frac{v|Y_{Qt}|}{|M_Q|}\ $ and
$\displaystyle\ \varepsilon_{Q_b}=\frac{v|Y_{Qb}|}{|M_Q|}\ $ and 
the relative phases $\theta_2^t$ and $\theta_2^b$ as
$\theta_2^t= Arg\left(\frac{Y_{qT}Y_{Qt} Y^t_2}{M_TM_QY^0_t}\right)$
and $\theta_2^b= Arg\left(\frac{Y_{qB}Y_{Qb}
  Y^b_2}{M_BM_QY^0_b}\right)$. Note that these 
perturbative expressions are only valid for $\varepsilon_i <
1$. Nevertheless they are very useful in identifying limits and
parameter behavior, and moreover the limit $\varepsilon_i <
1$ is the natural one as the top and bottom KK partners are
expected to be heavy enough to make the expansions converge. 
The first two terms of both expressions  always yield 
a suppression in the physical Yukawa coupling strength, irrespective of the phases
within the original fermion mass matrices. However, the third term,
proportional to $Y_2$, could induce an
overall enhancement of the top Yukawa coupling or of the bottom Yukawa
coupling, when the phases $\theta^{t,b}_2$ are such that
$-\pi/2<\theta^{t,b}_2<\pi/2$, and for sufficiently high values of
$Y_2$. An enhancement effect would be maximal when $\theta^{t,b}_2=0$.

Note that the top/bottom mirror sector,  even though essentially decoupled
from the light quarks, should still have some impact in the CKM quark
mixing matrix. In the same perturbative limit used to obtain Yukawa shifts, we can also
obtain an approximation to the corrections on $V_{tb}$, due to
the presence of the top/bottom vector-like mirrors. the value is shifted as
\bea
|V_{tb}| \simeq 1 - \frac{1}{2} |V_{cb}|^2 - \frac{1}{2} |V_{ub}|^2 -
\frac{1}{2} (\varepsilon_T-\varepsilon_B)^2 \, ,
\eea
where the first two terms represent the usual SM CKM unitarity
constraint, and the last term is the new contribution (where we
have eliminated the relative phases between $Y_{qT}$ and $M_T$, and
between $Y_{qB}$ and $M_B$ through a phase redefinition).
The current Tevatron and LHC average on $V_{tb}$, coming from single top
production is $V_{tb} = 1.009\pm 0.031$
\cite{Olive:2016xmw}, which gives a lowest bound of about $V_{tb} \sim
0.97$. That means that the corrections from our scenario
should be limited to about
\bea\label{vtbbound}
\frac{1}{2} (\varepsilon_T-\varepsilon_B)^2 \lsim (0.15)^2,
\eea
requiring that $Y v/M \lsim 0.15$ or $M \gsim 1$ TeV, unless a strong
cancellation between the top and bottom terms happens.
In a similar way, the rest of third row and third column CKM mixing angles $V_{ub}$, $V_{cb}$,
$V_{td}$ and $V_{ts}$ will receive corrections producing
deviations on the usual SM unitarity relations. For example we have
\bea
(1-\epsilon^2_{B})|V_{cb}|^2& \simeq& \left(1-|V_{cd}|^2-|V_{cs}|^2\right)
\eea
so that imposing the experimental uncertainties in $|V_{cd}|$,
$|V_{cs}|$ and $|V_{cb}|$ \cite{Olive:2016xmw}, we find that
\bea
\epsilon^2_{B}\lsim (0.44)^2.
\eea
This is a slightly less constraining bound on the vector-like sector,
compared to the one in Eq.~(\ref{vtbbound}). A thorough full fit
analysis on CKM unitarity is beyond the scope of this paper, although
should the $tth$ signal survive the higher luminosity data, with
improved constraints in the Higgs sector, such a study might become
useful.\footnote{Note that we are still assuming that first and second
  quark generations have highly suppressed Yukawa couplings with the
  top and bottom vector-like partners.} 

Finally, flavor mixing between vector-like quarks and the third
  generation can affect other flavor observables, particularly in
  $B$-physics. This was extensively discussed in the literature
  \cite{Bobeth:2016llm}, where a suppression of BR($B_s \to \mu {\bar \mu}$) and an enhancement in BR($B_d \to \mu {\bar \mu}$) are shown to be most likely. Here we will simply ask that the mixing in the
  bottom sector remains small. i.e. we should consider parameter space
points where the shift in bottom quark Yukawa coupling $\delta y_b$ is
small. Again, a full flavor analysis should be addressed if the $tth$ enhanced signal
is confirmed.

\subsection{Higgs Phenomenology}
\label{subsec:pheno}

\begin{figure}[t]
\center
\begin{center}
  \includegraphics[height=7.2cm]{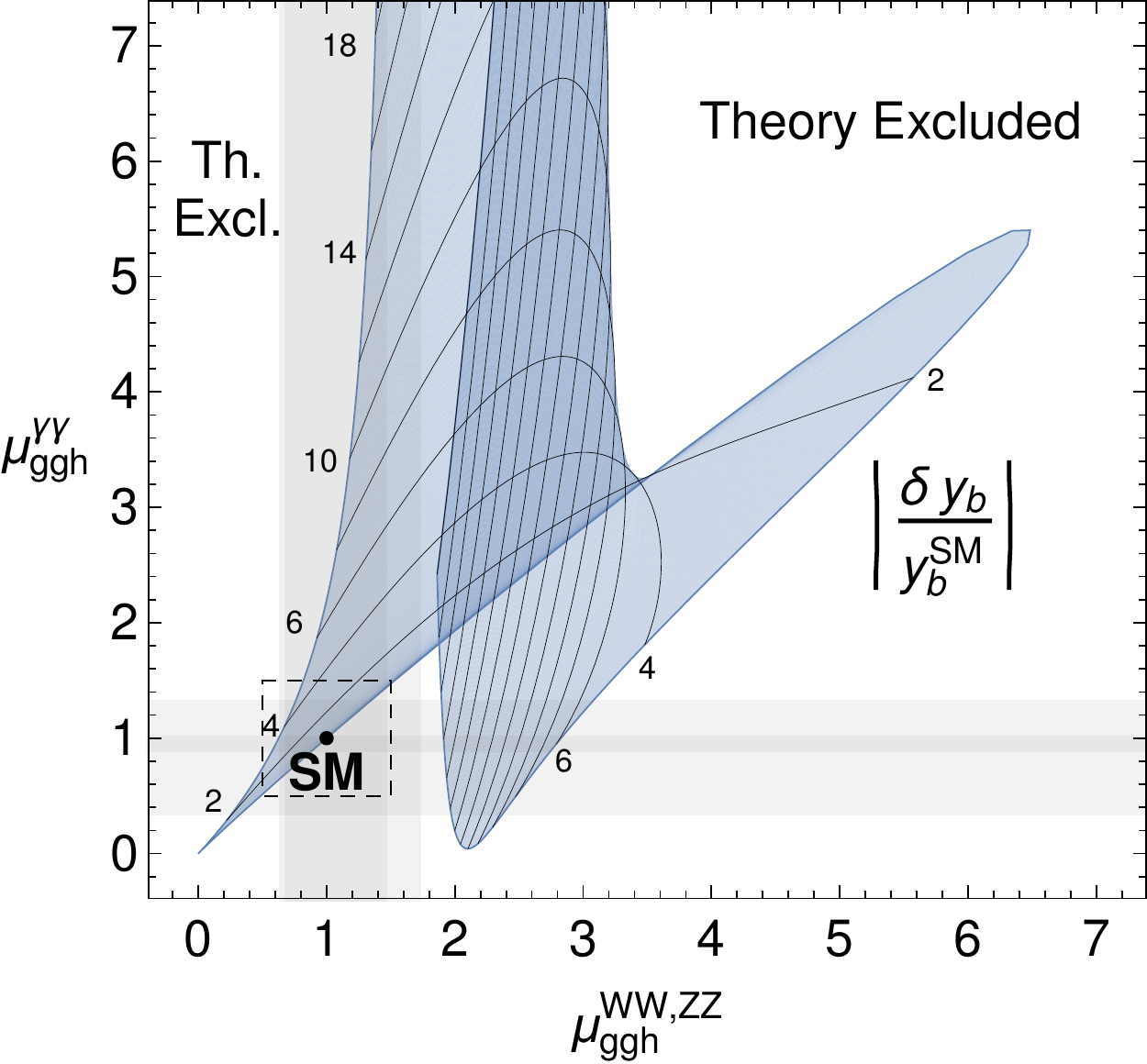}\hspace{.4cm}
  \includegraphics[height=7.2cm]{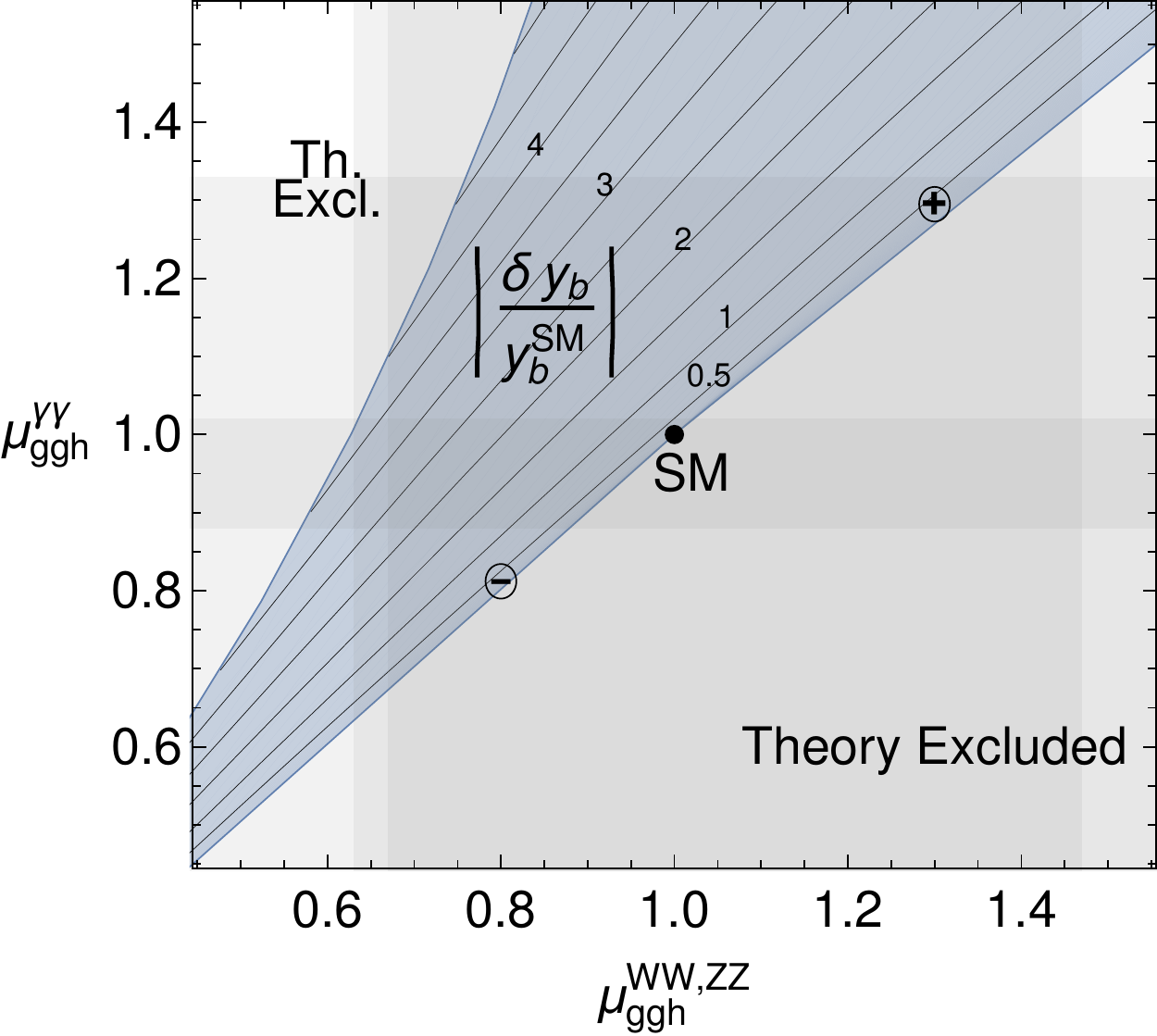}
 \end{center}
\vspace{.1cm}
\caption{Contours of the bottom quark Yukawa correction
  $\displaystyle\left|\frac{\delta y_b}{y_b^{SM}}\right|$ with respect to the gluon
  fusion  signal strengths $\mu_{ggh}^{VV}$ and
  $\mu^{\gamma\gamma}_{ggh}$. The right panel zooms in on the region
  marked by a dashed box in the left panel. The horizontal and
  vertical gray bands represent the experimental bounds set by the LHC
  RUN 1 (darker) and the preliminary data from LHC RUN 2 (lighter). The ``Theory
  Excluded'' regions are points excluded by the {\it Brane Higgs
    Limit} constraint.  Each contour is traced by varying the phase of
  $\delta y_b$ and we include two parameter space  points as example limits, marked by a 
 $\oplus$ and a $\ominus$, representing, respectively, an
  overall enhancement or suppression with respect to the SM predictions.
 }  \label{fig:ppb}
\vspace{.4cm}
\end{figure}

As we have seen earlier, the {\it Brane Higgs Limit} condition is
quite predictive, and easily falsifiable in the near future from LHC
Higgs data.
The first important point is that within our minimal general setup, all signal strengths associated with
the Higgs will deviate from the SM values {\it only} due to
shifts in the top and bottom quark Yukawa couplings.
This means that ratios of Higgs signal strengths involving electroweak
production processes,  and decays through the same  channels ``$ii$'', should be equal to one, i.e
\bea
\frac{\mu^{ii}_{VBF}}{\mu^{ii}_{Wh}} =\frac{\mu^{ii}_{VBF}}{\mu^{ii}_{Zh}}= 1
\eea
Also signal strengths involving decays into $WW$ should be equal to signals
with decays into $ZZ$, i.e.
\bea
\frac{\mu^{WW}_{ggh}}{\mu^{ZZ}_{ggh}}=\frac{\mu^{WW}_{tth}}{\mu^{ZZ}_{tth}}=\frac{\mu^{WW}_{Vh}}{\mu^{ZZ}_{Vh}}=\frac{\mu^{WW}_{VBF}}{\mu^{ZZ}_{VBF}}=1.
\eea
These are strong model dependent predictions, likely testable at the present RUN 2 at the LHC. 

Now, more specific to our setup, and as seen from Eqs. (\ref{Dtt})-(\ref{Dgaga}), the
corrections to all of the Higgs signal strengths depend only on four
parameters, i.e. the absolute values of the relative top 
and bottom Yukawa coupling deviations $|\delta y_t|$ and $|\delta y_b|$, and their
two phases. Moreover, only the $t{\bar t} h$ signal strengths depend on all
four parameters. We thus start exploring the dominant Higgs production mechanism,
the gluon fusion process, paying  particular attention to the signal strengths
$\mu^{\gamma\gamma}_{ggh}$ and $\mu^{WW,ZZ}_{ggh}$. These depend only
on the deviation of the bottom quark coupling (magnitude and phase).
It is therefore possible to study the relationship between these two
signal strengths, for different values of $\delta y_b$. This is plotted
in Fig. \ref{fig:ppb}, where we show that only a specific region in the
($\mu^{\gamma\gamma}_{ggh},\mu^{WW,ZZ}_{ggh})$ plane is allowed, due
to the {\it Brane Higgs Limit} constraint. The horizontal and vertical gray bands
correspond to limits set by LHC RUN 1 and preliminary LHC RUN 2 data, as summarized in Table \ref{tab:HIGGS_ttH_12}.
In the right panel of that figure, we zoom in the square enclosed by dashed lines in the left panel to consider signal strengths close
to the SM model value, and we can see that the region where
$\mu^{\gamma\gamma}_{ggh} < \mu^{WW,ZZ}_{ggh}$ is 
not allowed, thus providing a very simple and strong
prediction of the scenario. Corrections in the direction $\mu^{\gamma\gamma}_{ggh} > 
\mu^{WW,ZZ}_{ggh}$ are possible, but require increasingly large
deviations in the bottom Yukawa coupling. For relatively small values
of $\delta y_b$, one can still obtain important deviations in the
signals if one moves along the $\mu^{\gamma\gamma}_{ggh} \simeq \mu^{WW,ZZ}_{ggh}$ diagonal line. For
future use, we choose two points along that line, close to the boundaries
set by the LHC constraints. We denote them with a $\oplus$ and
$\ominus$, and they represent either an overall enhancement in the $ggh$ signal strengths,
or an overall suppression, with respect to the SM predictions.

\begin{figure}[t]
\center
\begin{center}
  \includegraphics[height=6.5cm]{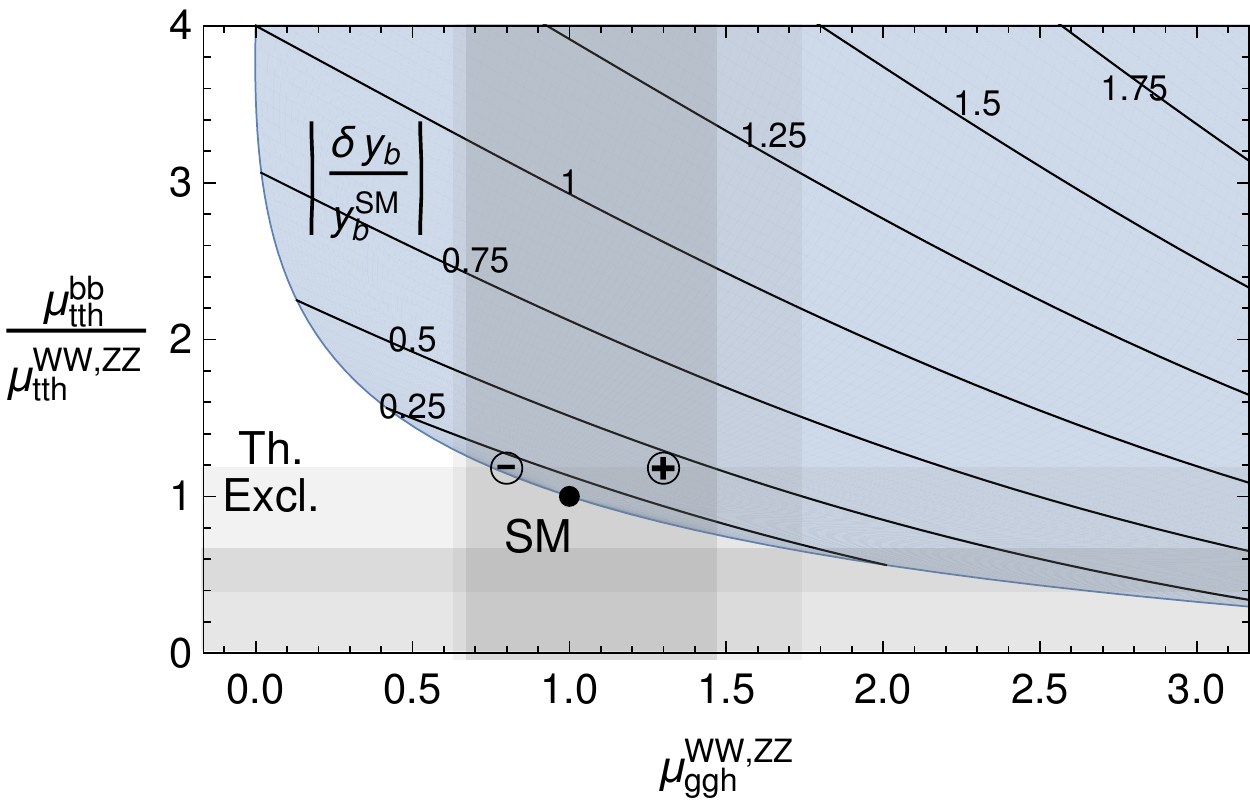}\hspace{.4cm}
 \end{center}
\vspace{.1cm}
\caption{Contours of the bottom quark Yukawa correction
  $\displaystyle\left|\frac{\delta y_b}{y_b^{SM}}\right|$ with respect to the gluon
  fusion  signal strength $\mu_{ggh}^{VV}$ and the ratio of the $t{\bar t}h$ signal
  strengths $\displaystyle\frac{\mu^{bb}_{tth}}{\mu_{tth}^{VV}}$. The horizontal and
  vertical grey bands represent the experimental bounds set by the LHC
  RUN 1 (darker) and the preliminary data from LHC RUN 2 (lighter). The ``Th. Excl.''
  region comprises all points excluded by the {\it Brane Higgs Limit} constraint.  Each contour is traced by
  varying the phase of $\delta y_b$ and we included two same parameter space
  points, marked by $\oplus$ and $\ominus$, as in the previous figure. 
 }  \label{fig:pptb}
\vspace{.4cm}
\end{figure}

Once the gluon fusion signals have been fixed, we can study the
effect on other signal strengths which receive corrections only through the bottom
quark Yukawa coupling. In particular we can explore how the ratios
$\displaystyle \frac{\mu^{bb}_{tth}}{\mu^{VV}_{tth}}
=\frac{\mu^{bb}_{Vh}}{\mu^{VV}_{Vh}}$ behave as a function of
$\mu^{VV}_{ggh}$ (all top quark Yukawa dependence cancels out in the
ratio). This is shown in Fig. \ref{fig:pptb}, where we
consider variations of the ratio
$\displaystyle\frac{\mu^{bb}_{tth}}{\mu^{VV}_{tth}}$ (with the corresponding LHC bounds
 represented by the horizontal gray bands), with respect to the gluon fusion
strength $\mu^{bb}_{tth}$. As we can see, the current experimental
data tend to prefer values for that ratio close to 1 or less,
therefore putting some pressure on the allowed parameter space. We can
see that if the $\mu^{bb}_{tth}$ signal strength is smaller than the $\mu^{VV}_{tth}$ one (both in $t{\bar t}h$
production), then the data prefers a slight enhancement in the gluon
fusion production strength. Conversely, if the $\mu^{bb}_{tth}$ signal is enhanced, then gluon fusion
signals should be suppressed. Overall, the deviations on the bottom
quark Yukawa coupling must be kept small, unless the $\mu^{bb}_{tth}$ signal
happens to be very much larger than the $\mu^{VV}_{tth}$ signal. The chosen example  points $\oplus$ and
$\ominus$ stay within a ratio of $\mu_{tth}$ production signals close to 1.

\begin{figure}[t]
\center
\begin{center}
  \includegraphics[width=5.32cm,height=8cm]{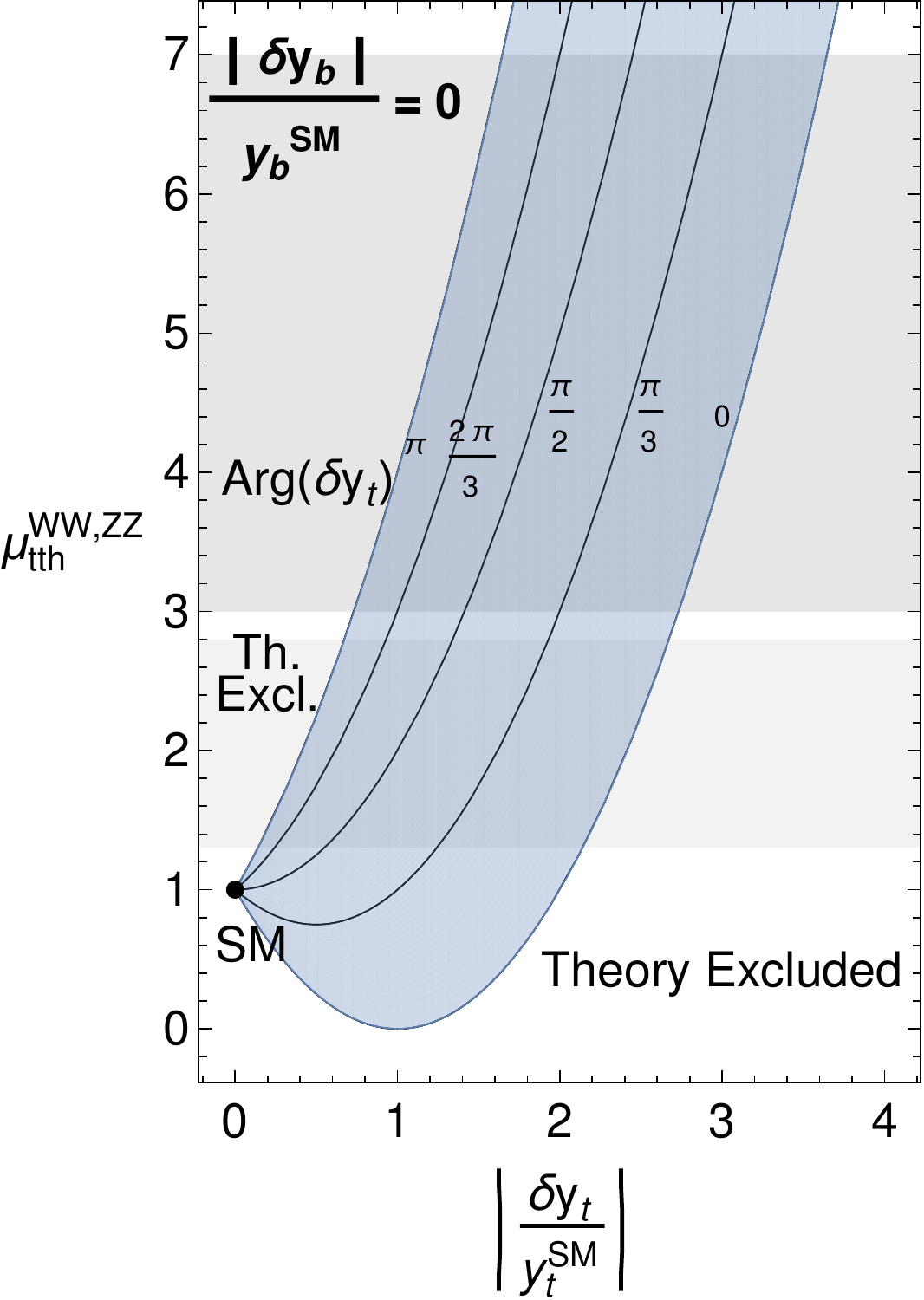}\hspace{.1cm}
  \includegraphics[width=5.32cm,height=8cm]{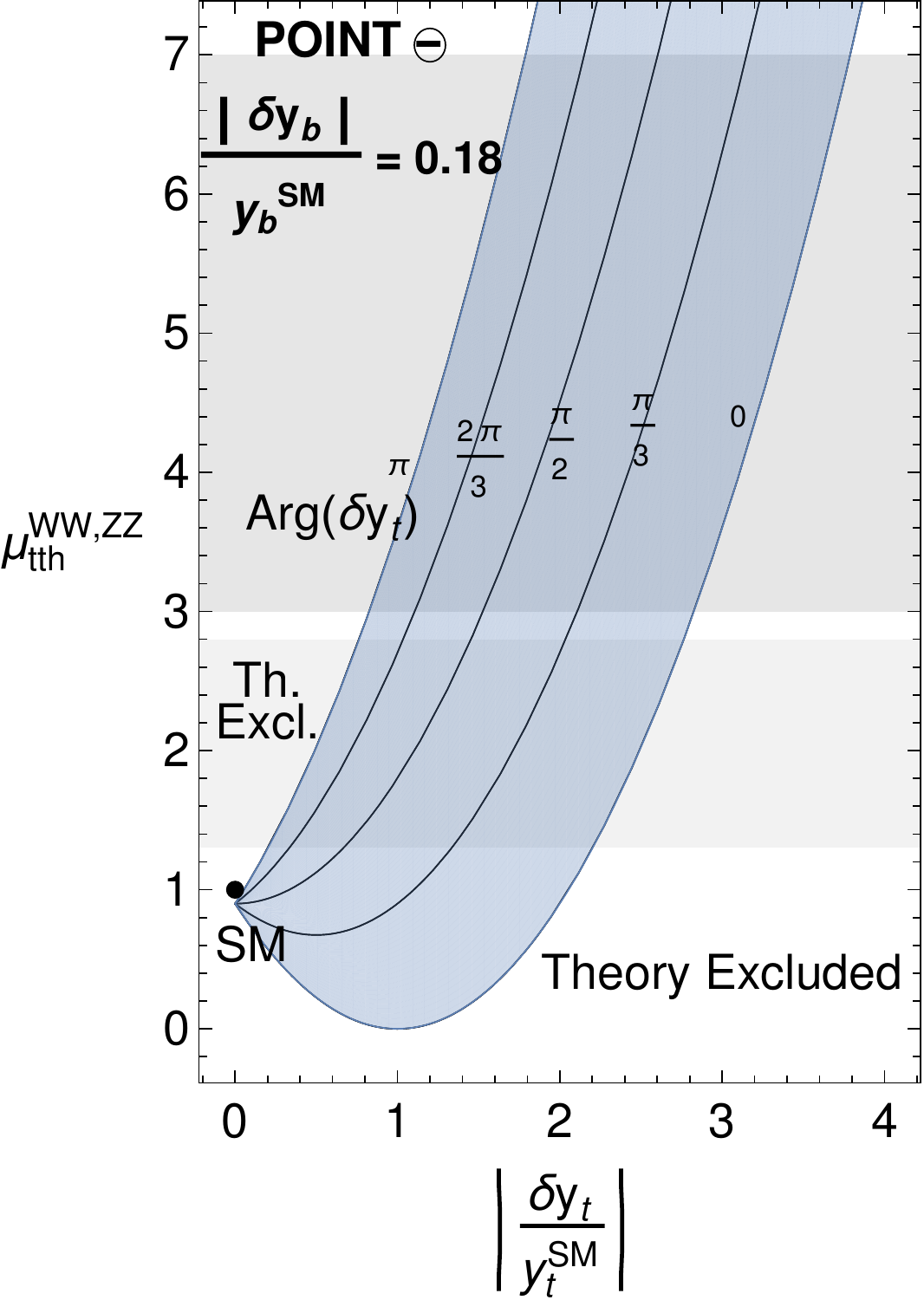}\hspace{.1cm}
  \includegraphics[width=5.32cm,height=8cm]{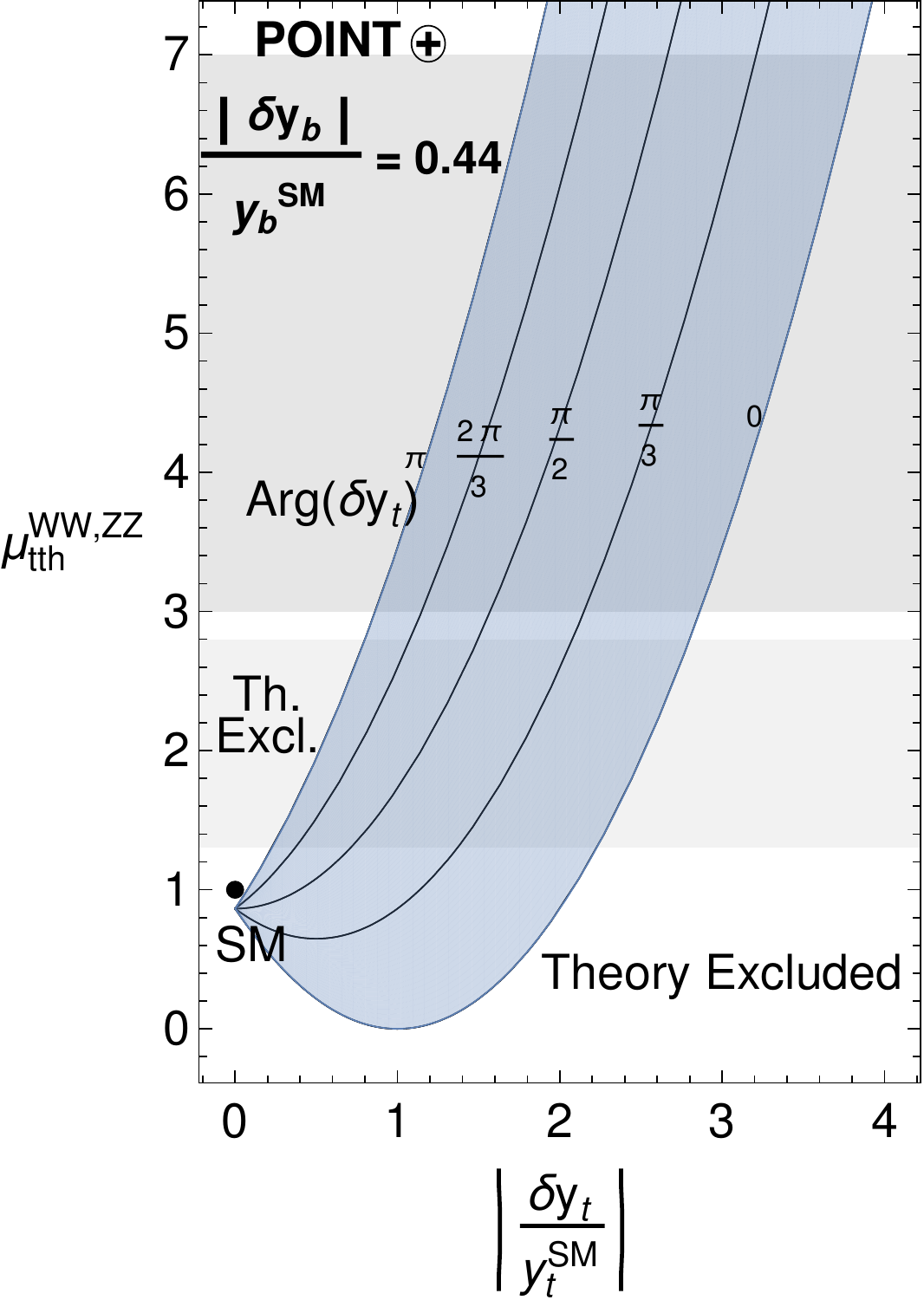}
 \end{center}
\vspace{.1cm}
\caption{
Higgs signal strength $\mu_{tth}^{VV}$ with
  respect to the top quark Yukawa coupling  correction $\displaystyle \left|\frac{\delta
    y_t}{y_t^{SM}}\right|$. The contours trace points with constant
  value of the phase of $\delta y_t$, and the horizontal gray bands
  represent the experimental bounds set by the LHC RUN 1 (darker) and the
  preliminary data from LHC RUN 2 (lighter). The ``Theory Excluded'' regions are
  excluded by the {\it Brane Higgs Limit} constraint.
  In the left panel, we consider a parameter space point where the
  bottom Yukawa coupling is SM-like. In the central panel the bottom
  quark Yukawa coupling is corrected by 18\% (point $\ominus$) and
  the right panel has a bottom quark Yukawa correction of 44\% (point $\oplus$).  }  \label{fig:pptV}
\vspace{.4cm}
\end{figure}

Once we analyzed the restriction on the deviations from bottom quark Yukawa couplings, we can investigate the signals that do depend on the top Yukawa
coupling deviations. In Fig. \ref{fig:pptV} we choose to study the variation of $\mu^{VV}_{tth}$ with respect to 
the top Yukawa deviation $|\delta y_t|$. The rest of $t{\bar t}h$ signals
strengths can be obtained from ratios of other Higgs production
signals strengths, since for example $\displaystyle \mu^{\gamma\gamma}_{tth}
=\frac{\mu^{\gamma\gamma}_{ggh}}{\mu^{VV}_{ggh}}\ \mu^{VV}_{tth}$. We fix the values of the bottom
quark Yukawa coupling in three limits, i.e. when $y_b$ is SM-like
($\delta y_b =0$), when it has a 18\% correction
($\left|\delta y_b/y_b^{SM}\right| =0.18$, corresponding to the point $\ominus$), and when
it has a 44\% correction ($\left|\delta y_b/y_b^{SM}\right|
=0.44$, corresponding to the point $\oplus$).
As can be seen in Fig. \ref{fig:pptV}, 
for moderate values of $\delta y_b$ there is very mild dependence on $\delta y_b$, so that the three
panels show very similar behavior of the signal strength as a function of the deviations in top quark Yukawa couplings.  The parameter space region is a diagonal
band, and we show contours of the phase of the top Yukawa shift $\delta
y_t$, tracing the band diagonally. The dependence is very sensitive to
variations in the phase of the shift of the top Yukawa coupling. We
can clearly see that if the magnitude of the top Yukawa deviation is
less than 1 (the natural expectation for heavy KK top partners),  in
order to obtain a signal enhancement (as hinted by LHC data), the
phase must be close to $\pi$. This is in agreement with the perturbative
expressions obtained earlier for the Yukawa shifts and it corresponds
to values of the mass matrix phase $\theta^t_2$ close to 
0.


\section{``Brane Higgs Limit'' in Randall Sundrum models}
\label{sec:RS}

In this section we describe briefly how to reproduce the previous
phenomenological scenario within the context of the Randall Sundrum model \cite{RS1}.
Consider a sector of a 5D scenario with a 5D top quark,
i.e. a doublet fermion $Q(x,y)$ and a singlet $T(x,y)$ defined by the following action:
\bea
&&S=-i\int d^4xdy \sqrt{-g}\left[\bar{Q} D\!\!\!\!/ Q + c_q
  \sigma'\bar{Q}Q + \bar{T} D\!\!\!\!/ T + c_t  \sigma'\bar{T}T
  \right. \non\\
 && \left.+ \delta(y-y_1) \left(\alpha_{qL} \bar{Q}_L \partial \!\!\!/ Q_L +
  \alpha_{qR} \bar{Q}_R \partial\!\!\!/ Q_R + \alpha_{tL}
  \bar{T}_L \partial \!\!\!/ T_L +
  \alpha_{tR} \bar{T}_R \partial \!\!\!/ T_R   \right)\right. \non\\
  && \left.+ \delta(y-y_1) \left( Y_1 H \bar{Q}_L T_R + Y_2 H \bar{Q}_R T_L +h.c.  \right)\right]
\label{5Daction}
\eea
where  $D\!\!\!\!/ = \gamma_A e^M_A D_M$ and $\partial \!\!\!/ =
\gamma_a e^\mu_a \partial_\mu$, with $\gamma_A$ the 5D
gamma matrices, $e^M_A$ the vielbein, and
$D_M=(\partial_M+\Gamma_M) $  the 5D covariant derivative involving
the spin connection $\Gamma_M$, with $\Gamma_\mu=\frac{1}{2}
\gamma_5\gamma_\mu \sigma'$ and $\Gamma_5=0$.
The fifth dimension is understood as an interval, with the
boundary terms fixing the boundary conditions of the fields.
We have added a set of fermion kinetic terms localized at the boundary
$y=y_1$. Other boundary fermion kinetic terms, involving
$y$-derivatives, are allowed but we leave them out for
simplicity.  Also note that we should only consider positive brane kinetic
term coefficients $\alpha_i$, in order to avoid tachyons and/or ghosts
\cite{delAguila:2006atw,Carena:2004zn}.

We also consider Higgs localized Yukawa couplings on the same boundary. Note that  the doublet $Q$
is vector-like in 5D, and we  define $Q=Q_L+Q_R$ and $T=T_L+T_R$
where $Q_L, T_L$ and $Q_R,T_R$ are the left and right handed components.  

The background spacetime metric is assumed to take the form 
\bea
ds^2=e^{-2\sigma(y)}\eta_{\mu\nu} dx^\mu dx^\nu + dy^2\label{warpedmetric}
\eea
where $\sigma(y)=ky$ is known as the warp factor (note that the
signature is $(-,+,+,+,+)$) and with $k\sim M_{Pl}$ being the 5D curvature. 
We assume that $\sigma(y_0)=1$ and $\sigma(y_1)\simeq 34$ such that
there are some 15 orders of magnitude of scale hierarchy between both boundaries.

In the absence of fermion brane kinetic terms (proportional to $\alpha_i$'s), this
setup produces a tower of Kaluza Klein (KK) modes, such that the lowest lying
modes of the doublet and singlet fields have wavefunctions
exponentially localized towards either of the boundaries \cite{Davoudiasl:1999tf}. The
localization depends on the value of the 5D fermion mass parameters
$c_q$ and $c_t$. When $c_q < 1/2$ and $c_t > -1/2$,  the zero
modes of $Q$ and $T$ will be localized near the $y=y_1$ boundary, and
will be identified as the two chiral components of the SM top quark. The rest of SM
quarks will be obtained in a similar way, but the value of their bulk mass will localize
them towards the $y=y_0$ boundary. Because the Higgs boson  is by
construction located at the $y=y_1$ boundary, the top quark will be
``naturally'' heavy (coupled strongly 
to the Higgs) whereas the rest of quarks quarks are lighter, since
they couple to the Higgs weakly due to their geographical separation.
On the other hand, the excited modes of all fermions will be very
heavy and localized towards the $y=y_1$ boundary; their typical KK
masses are of order $M_{\rm Pl} e^{-k y_1} \sim TeV$ and they will also
couple strongly with the Higgs.

The scenario that we call the {\it Brane Higgs Limit}, requires the
presence of $only$ the top quark and bottom quark heavy partners, which means the rest of KK 
partners should be decoupled (i.e. much heavier). For this we turn-on the fermion brane kinetic
terms (the $\alpha_i$'s) of the 5D top and bottom quarks.  There will still be massless fermion modes
(associated to the SM top and bottom quarks\footnote{Which acquire
  their SM masses after electroweak symmetry breaking, like in the SM.}), but it now becomes possible to 
lower the masses of the KK top and bottom modes. In general, it is possible to obtain
analytically the associated KK spectrum (before electroweak symmetry
breaking) in terms of Bessel functions. Nevertheless, since we are
mainly interested in the top quark, it is much simpler and transparent to
treat the special case where $c_q=c_t=0$. These simple bulk masses are
perfectly top-like, and they have the advantage of producing very
simple equations of motion. 
The usual dimensional reduction procedure involves a mixed separation
of variables performed on the 5D fermions, i.e.
\bea
Q_L(x,y)= Q_L(y) t_L(x)\\
Q_R(x,y)= Q_R(y) t_R(x)\\
T_L(x,y)= T_L(y) t_L(x)\\
T_R(x,y)= T_R(y) t_R(x)
\eea
where $t_L(x)$ and $t_R(x)$ are the left and right handed components
of 4D fermions (the lightest of which is the SM top quark). From
there, one must solve for the KK profiles along the extra dimension 
$Q_L(y)$, $Q_R(y)$, $T_L(y)$ and $T_R(y)$. In the simple case of
$c_q=c_t=0$, and before electroweak symmetry breaking,the equation
for the profile $\tilde{Q}_R(y)= e^{-2\sigma(y)} Q_R(y)$, for example,  becomes
\bea
\left(e^{- k y} \tilde{Q}'_R\right)'+
m^2 e^{k y} \tilde{Q}_R =0
\eea
with Dirichlet boundary condition on the $y=0$ boundary (since there
are no kinetic terms there). The solution is simple,
\bea
\tilde{Q}_R(y)= N_Q \sin{\left(\frac{m(e^{k y}-1)}{k} \right)} 
\eea
which obviously vanishes at $y=0$.
The brane kinetic term on the $y=y_1$ boundary  enforces a matching
boundary condition at that location, and that boundary condition fixes the
spectrum of the whole tower of KK modes. In this simple case, the KK
spectrum of the 5D fermions $Q(x,y)$ and $T(x,y)$ is given by 
\bea
\tan{\left(
  \frac{m e^{ky_1}}{k}\right)} =-
\sqrt{\frac{\alpha_{qL}}{\alpha_{qR}}} \tan{\left(
  m e^{ky_1}\sqrt{\alpha_{qR} \alpha_{qL}} \right)}
\eea
and 
\bea
\tan{\left(
  \frac{m e^{ky_1}}{k}\right)} =-
\sqrt{\frac{\alpha_{tL}}{\alpha_{tR}}} \tan{\left(
  m e^{ky_1}\sqrt{\alpha_{tR} \alpha_{tL}} \right)},
\eea
in agreement with the flat metric limit considered in
\cite{delAguila:2006atw}. 
With a further simplification, taking
$\alpha_{qR}=\alpha_{qL}=\alpha_q$ and
$\alpha_{tR}=\alpha_{tL}=\alpha_t$, the conditions become 
\bea
\tan{\left(
  \frac{m e^{ky_1}}{k}\right)} = - \tan{\left(
  m e^{ky_1}\alpha_i \right)}
\eea
with a spectrum given by
\bea
m_n= \frac{n \pi }{1+k \alpha_i} k e^{-ky_1}
\eea
for $n=0,1,2,3...$ This shows that, indeed, the spectrum of the KK
tops can be significantly reduced in the presence of brane kinetic
terms.

In the scenario we have in mind, only the 5D top and bottom quarks have large brane
kinetic terms (without further justification) and therefore their
associated KK modes can be much lighter than the rest, maybe as light
as 1 TeV. At the same time, the rest of quarks and KK gauge bosons
follow the usual RS pattern with KK 
masses maybe an order of magnitude larger ($\sim 10$ TeV). In this
limit, flavor and precision electroweak bounds are much safer and the
main phenomenological effects of the model may occur within the 
Higgs sector of the scenario. 

If we decouple the $up$, $down$, $strange$ and $charm$ heavy KK
quarks, the fermion mass matrices will involve only SM quarks along
with KK $tops$ and KK $bottoms$. Mixing between light quarks
localized near $y=0$ and heavier quarks localized near $y=y_1$ is
going to be CKM suppressed, as usual in RS, and therefore the mass
matrices to consider have the same form of those in Eqs.~(\ref{Mt})
and (\ref{Mb}), but with the phases $\theta_1^t =\theta^b_1=0$. Indeed,
the values of the off-diagonal terms in the mass 
matrices are now associated to the 5D Yukawa interactions localized at
the $y=y_1$ brane and in this case are such that 
\bea
Y_t^0&=& Y_{33}^{5D} f_{q_L} f_{t_R}\\
(Y_1)_{nm}  &=& Y_{33}^{5D} f_{Q^n_L} f_{T^m_R}\\
(Y_{qT})_n&=&Y_{33}^{5D} f_{q_L} f_{T^n_R}\\
(Y_{Qt})_n&=&Y_{33}^{5D} f_{Q^n_L} f_{t_R}
\eea
where $Y_{33}^{5D}$ is the 5D top Yukawa coupling, and where the $f_i$'s are
the wavefunctions evaluated at the $y=y_1$ brane\footnote{Note that
  another effect of the brane kinetic terms is to suppress the value of
  wavefunctions through normalization, due to the new brane localized kinetic
  terms. Nevertheless,  in order to obtain the top quark 
  mass, the 5D coupling $Y_{33}$ must be enhanced accordingly
  and thus the wavefunction suppression is compensated by a coupling
  enhancement, while remaining in a perturbative regime
  \cite{Carena:2004zn}.}, with $q_L$ and 
$t_R$ being zero modes and $Q^n_L$  and $T^n_R$ representing the $n^{th}$ KK modes.
We can see that all these terms share the same phase, so that we can
set $\theta_1^t =\theta^b_1=0$.

Now, consider first a mass matrix with only one KK level ($n=1$), so that the matrices are exactly
the same as before and thus the corresponding effect in Higgs production will
come from the sum   
\bea
\sum_n \left({\frac{y^u_{nn}}{m^u_{n}}}\right)&=&
\frac{1}{v} \frac{1 + 3 \varepsilon_{Q_t} \varepsilon_T
  \frac{|Y_2|}{|Y_t^0|}e^{i\theta^t_2}\left(1
  -\frac{|Y_1||Y_t^0|}{|Y_{Qt}||Y_{qT}|}
  \right)}{1 + \varepsilon_{Q_t} \varepsilon_T
  \frac{|Y_2|}{|Y_t^0|}e^{i\theta^t_2}\left(1
  -\frac{|Y_1||Y_t^0|}{|Y_{Qt}||Y_{qT}|}
  \right)}.
\eea
It becomes then apparent that $\left(1
-\frac{|Y_1||Y_t^0|}{|Y_{Qt}||Y_{qT}|}\right)=0$ due to the structure of the
5D couplings. It is important  that the zero modes (SM top and
bottom) come from the same 5D fermion as the KK modes, since the cancellation will
only happen if they all share the  same 5D Yukawa coupling.
It turns out that it is simple to prove that if we take into
account the complete towers of KK tops we still have
\bea
\sum^\infty_n \left({\frac{y^u_{nn}}{m^u_{n}}}\right)&=&\frac{1}{v}
\eea
 and similarly for the bottom quarks, and thus the Higgs phenomenology of
 this scenario is indeed the same as in the {\it Brane Higgs limit} introduced
 earlier in a bottom-up approach, since relative corrections due to the KK
 modes of the up, down, strange and charm quarks, will scale as
 $\left(m_{KK}^{tops}/m_{KK}^{rest}\right)^2 \sim  1\%$ (assuming that
 the rest of KK quarks are an order of magnitude heavier than KK tops/bottoms). 
 Note that if the Higgs boson is not exactly localized at the boundary, then
 the cancellation will not be exact and new corrections will arise.

The contributions of this RS scenario to flavor and precision electroweak
observables will be limited to effects due to the  mixing of top and
bottom with the vector-like partners, since  we are considering very
heavy KK gauge bosons ($\sim 10$ TeV). As pointed out earlier, Yukawa coupling mixing
effects can lead to deviations in $|V_{tb}|$ which can easily be kept under
control. Also effects can appear in the couplings $Z \to
b_R{\bar b}_R$  and in $Z \to b_L{\bar b}_L$ \cite{Agashe:2006at}, but
since we consider the contribution from heavy KK gauge
bosons to be suppressed, only Yukawa coupling mixings contribute, limiting the correction.
In the usual RS scenario, it was already possible to find points in
the $(c_b,c_{q_3})$ plane, such that $Zb{\bar b}$ couplings remain within
experimental bounds, with all the SM masses and mixings correctly
obtained \cite{Casagrande:2008hr}. In our scenario,  finding
parameter points safe from precision tests will be even easier, since the source of corrections is further limited. 


\section{Conclusion}
\label{sec:conclusion}


In this work, we presented a simple explanation of the possible
enhancement in the $t {\bar t} h$ associated production seen at the
LHC. We added one $SU(2)$ doublet, and two $SU(2)$ singlet vector-like
quarks to the matter content of the SM, as partners of the third
family, and allow significant mixing between these with the third
family only. After electroweak symmetry breaking, Yukawa couplings
induce off-diagonal terms into the fermion mass matrix and, once in the
physical basis, the top and bottom Yukawa couplings and their corresponding
masses loose their SM alignment.    
With the proper sign (or phase), this misalignment can induce an
enhancement of the top quark Yukawa coupling, and thus increase the
cross section for  $t {\bar t}h$ production. But the mechanism should
also affect other observables in the Higgs sector, in particular, the
cross section for Higgs production through gluon fusion. This is the
main production channel for the SM Higgs and, being a radiative effect,
it receives a contribution from all the fermions in the model. Each
contribution is proportional to the ratio of the Yukawa fermion
couplings to the fermion mass, so that the main contributions come 
from the top quark (with an enhanced Yukawa coupling), and from the 
new vector-like quarks. 

We showed that working in a particular limit of parameter space, the
corrections to gluon fusion caused by the top Yukawa coupling
enhancement are exactly offset by the contributions of the new top
partners, so that the overall top sector of our scenario (top quark
plus heavy partners) gives the same contribution as the single top quark
contribution in the SM. We call this scenario the  {\it Brane
  Higgs limit} and it yields extremely predictive relationships
between the productions cross sections ($ gg $ and $t {\bar t}h$) and
decay branching ratios for the Higgs bosons (into  $WW, \, ZZ, \,
b{\bar b}$ and  $\gamma \gamma$), 
where the only free parameters are the absolute values of the shifts
in the Yukawa couplings of the top and the bottom quarks, and their
phases. For instance,  in $t {\bar t}h$ production, if the branching
ratio into $b {\bar b}$ is smaller than that into $VV$, then the gluon
fusion production cross section must be also greater than its
SM value, and conversely, an enhancement of the  $b {\bar b}$
branching ratio in $t {\bar t}h$ production indicates a suppressed
gluon fusion signal. Overall, the deviations in the Yukawa coupling of
the bottom quark are constrained to be small, unless new data
indicates a significant enhancement of the $b {\bar b}$ branching
ratio. The scenario we consider predicts that any enhancement or
suppression in the $\gamma \gamma$ signal should be matched with
identical enhancement or suppression in $VV=WW, ZZ$ decays (for gluon fusion
production), or at least remain always slightly higher than decays into $VV=WW,ZZ$,
but never lower. Finally, a shift in the top quark Yukawa coupling
will affect all $t {\bar t}h$ signals through the production cross section, and
enhancement or suppression will depend on the phase of the shift.   

The mixing in the top quark and bottom quark sectors should have also
consequences in the Cabibbo-Kobayashi-Makskawa (CKM) mixing
matrix $V_{CKM}$, as well as in precise electroweak measurements. We
briefly discussed how the scenario affects (and thus is constrained
by) the $V_{tb}$ entry of $V_{CKM}$ and the decay $Z \to b {\bar b}~(A_{FB}^b)$.  

Finally we showed that the phenomenology we described here depends on a
specific structure of the fermion mixing matrix, mixing top quark with
its partners. In particular the Yukawa coupling matrix should have
a vanishing determinant, and thus some mechanism or flavor symmetry
should be invoked to realize the scenario. The
required structure is  naturally realized in a Randall Sundrum model
without a need for flavor symmetries. A key ingredient of this scenario is the
presence of brane kinetic terms for the top and bottom, which can then
result in lighter $n=1$ KK modes for the top and bottom partners, but
heavy masses for all other KK modes. If the Higgs is localized exactly
at the boundary, the overall phenomenology of the simple model
introduced here is essentially recovered (i.e the cancellation of the terms happening in the
gluon fusion calculation occurs by construction, even if in this case
the effect comes from a complete tower of KK states). 

The  model presented here thus has a simple theoretical
realization, is highly predictable, and can be tested (or ruled out) by more
precise measurements of the Higgs signal strength in RUN 2 at LHC.


\section{Acknowledgments}


M.T. would like to thank FRQNT for partial financial support under
grant number PRCC-191578 and M.F. acknowledges NSERC  support under
grant number SAP105354.

\end{document}